\documentclass[aps,pre,groupedaddress,amsmath,showkeys]{revtex4}
    \usepackage{bm}
    \usepackage{graphicx}
\newcommand\beq{\begin{equation}}
\newcommand\eeq{\end{equation}}
\newcommand\beqa{\begin{eqnarray}}
\newcommand\eeqa{\end{eqnarray}}
\newcommand{\dd}{\text{d}}
\newcommand{\nn}{\nonumber\\}

\newcommand{\ee}{\text{e}}

\newcommand{\al}{\alpha}
\newcommand{\ati}{\widetilde{a}}

\newcommand{\ellmin}{\ell_{\text{min}}}
\newcommand{\nutz}{\zeta}
\newcommand{\nuzt}{\nu_{0|2}}
\newcommand{\nuto}{\nu_{2|1}}
\newcommand{\nuzth}{\nu_{0|3}}
\newcommand{\nufz}{\nu_{4|0}}
\newcommand{\nutt}{\nu_{2|2}}
\newcommand{\nuzf}{\nu_{0|4}}

\newcommand{\omegazt}{\omega_{0|2}}
\newcommand{\omegato}{\omega_{2|1}}
\newcommand{\omegazth}{\omega_{0|3}}
\newcommand{\omegafz}{\omega_{4|0}}
\newcommand{\omegatt}{\omega_{2|2}}
\newcommand{\omegazf}{\omega_{0|4}}
\newcommand{\nubar}{\omega_{0|2}}

\newcommand{\omeganu}{\nu}

\begin{document}
\title{Simple shear flow in inelastic Maxwell models}

\author{Andr\'es Santos}
\email{andres@unex.es}
\homepage{http://www.unex.es/eweb/fisteor/andres/}
\author{Vicente Garz\'o}
\email{vicenteg@unex.es}
\homepage{http://www.unex.es/eweb/fisteor/vicente/}
\affiliation{Departamento
de F\'{\i}sica, Universidad de Extremadura, E-06071 Badajoz, Spain}
\date{\today}

\begin{abstract}
The Boltzmann equation for  inelastic Maxwell models  is considered
to determine the velocity moments through fourth degree in the
simple shear flow state. First, the rheological properties (which
are related to the second-degree velocity moments) are {\em exactly}
evaluated in terms of the coefficient of restitution $\alpha$ and
the (reduced) shear rate $a^*$. For a given value of $\alpha$, the
above transport properties decrease with increasing  shear rate.
Moreover, as expected, the third-degree and the asymmetric
fourth-degree moments vanish in the long time limit when they are
scaled with the thermal speed. On the other hand, as in the case of
 elastic collisions, our results show that, for a given value of
$\alpha$, the scaled symmetric  fourth-degree moments diverge in
time for shear rates larger than a certain critical value
$a_c^*(\alpha)$ which decreases with increasing dissipation. The
explicit shear-rate dependence of the fourth-degree moments below
this critical value is also obtained.
\end{abstract}

\maketitle

\maketitle

\section{introduction}

On of the most widely studied inhomogeneous states in granular gases
is the so-called simple or uniform shear flow (USF) \cite{C90,Go03}.
This state is characterized by a constant density $n$, a uniform
granular temperature $T$, and a linear velocity profile $u_x=a y$,
where $a$ is the constant shear rate. The temperature changes in
time due to two competing effects: the viscous heating and the
inelastic collisional cooling. Depending on the initial condition,
one of the effects prevails over the other one so that the
temperature either increases or decreases in time, until a steady
state is reached for sufficiently long times. After a short kinetic
regime, the time evolution and the steady state of the system admits
a non-Newtonian hydrodynamic description \cite{SGD04,AS07}
characterized by  shear-rate dependent  viscosity and normal stress
differences.

The prototypical model of granular gases consists of inelastic hard
spheres  (IHS) with a constant coefficient of normal restitution
$\alpha \leq 1$. In the low-density limit, all the relevant
information on the system is provided by the one-particle velocity
distribution function $f(\mathbf{r},\mathbf{v};t)$, which  obeys the
Boltzmann equation  \cite{BP04}. However,  it is generally not
possible to get exact analytical results from the Boltzmann equation
for IHS, especially in far from equilibrium situations such as the
USF. Consequently, most of the  analytical results reported in the
literature have been derived by using approximations and/or kinetic
models
\cite{LSJC84,JR85,JR88,LB94,SGN96,BRM97,CR98,MGSB99,MG02,L04,L06,G06}.

The lack of  exact analytical results can be overcome in some
situations by considering the so-called inelastic Maxwell models
(IMM), where the collision rate is independent of the relative
velocity of the two colliding particles. These models  have received
a lot of attention in the last few years since they allow for the
derivation of a number of exact results
\cite{BCG00,CCG00,NK00,C01,EB02a,EB02,EB02bis,BMP02,KN02,KN02bis,NK02,BC02,MP02,MP02bis,BC03,BCT03,SE03,S03,NK03,
G03,BE04,GA05,BG06,ETB06,BTE07,S07,GS07}. Therefore,  the influence
of inelasticity on the dynamic properties can be analyzed  without
introducing additional, and sometimes uncontrolled, approximations.
In addition,  it is interesting to remark that recent experiments
\cite{KSSAON05} for magnetic grains with dipolar interactions turn
out to be well described by IMM.

In the context of the USF, the rheological properties, which are
related with the second-degree velocity moments, have been obtained
exactly in the steady state from the Boltzmann equation for IMM
\cite{C01,G03}. However, even though these properties are physically
important, they provide a partial piece of information about the
velocity distribution function $f$, especially in the high-velocity
region, where higher degree velocity moments play a prominent role.
By symmetry reasons, the third-degree moments are expected to vanish
in the USF. Therefore, the first non-trivial moments beyond the ones
associated with the rheological properties are the fourth-degree
moments. Their knowledge provides relevant information about the
combined effect of shearing and inelasticity on the velocity
distribution.

The effort of going from second-degree to fourth-degree moments in
the USF problem can be justified by a number of reasons. For
instance, their knowledge is needed to evaluate  transport
properties in situations slightly perturbed from the USF state
\cite{G07}, which allows one to perform a linear stability analysis
around that state. Another interesting issue is to explore whether
or not the divergence of the fourth-degree moments for elastic
Maxwell molecules beyond a certain critical shear rate
\cite{GS03,SGBD94,SG95} is also present in the inelastic case and,
if so, to what extent.

The main aim of this paper is to determine the fourth-degree moments
of IMM subject to USF. This can be carried out thanks to recent
derivations by the authors of the fourth-degree collisional moments
for IMM \cite{GS07}. Those moments are proportional to an effective
collision frequency $\nu_0$, which in principle can be freely
chosen. Here we will consider two classes of IMM: (a) a collision
frequency $\nu_0$ independent of temperature (Model A) and (b) a
collision frequency $\nu_0(T)$ monotonically increasing with
temperature (Model B). While Model A is closer to the original model
of Maxwell molecules for elastic gases \cite{GS03,TM80}, Model B,
with $\nu_0(T)\propto T^{1/2}$, is closer to IHS. The possibility of
having a general function $\nu(T)$ is akin to the class of inelastic
repulsive models recently introduced by Ernst and co-workers
\cite{ETB06,BTE07}. As will be shown below, Model A and B yield the
same results in the steady state. In particular, the reduced shear
rate $a^*=a/\nu_0$ in the steady state is a ``universal''
well-defined function  $a_s^*(\alpha)$ of the coefficient of
restitution $\alpha$. The main feature of Model A is that $a^*$ does
not change in time and so a steady state does not exist, except for
the specific value $a^*=a_s^*(\alpha)$. However, a non-Newtonian
hydrodynamic regime (in which $a^*$ and $\alpha$ are independent and
arbitrary parameters) is reached for asymptotically long times. This
allows to study analytically the combined effect of both control
parameters on the (scaled) velocity moments for Model A.

The plan of the paper is as follows. In Section \ref{sec2}, the
Boltzmann equation for IMM is introduced and the explicit
expressions for the collisional moments through fourth-degree are
given. Section \ref{sec3} deals with the rheological properties (a
nonlinear shear viscosity $\eta^*$ and a viscometric function
$\Psi$) of the USF state, which are related to the second-degree
velocity moments (pressure tensor). While Model A lends itself to
obtain the {\em exact} forms of $\eta^*$ and $\Psi$ as nonlinear
functions of $a^*$ and $\alpha$, that is not the case for Model B
since those rheological quantities require to be solved numerically,
except in the steady state. For elastic collisions ($\alpha=1$), our
expressions of $\eta^*$ and $\Psi$ obtained for Model A reduce to
the results derived  long time ago by Ikenberry and Truesdell
\cite{IT56} for Maxwell molecules. The third- and fourth-degree
moments for Model A are analyzed in Section \ref{sec4}. As expected,
the results show that, when the third- and asymmetric fourth-degree
moments are conveniently scaled with the thermal speed, they vanish
in the long time limit. This is not the case for the symmetric
fourth-degree moments. In a  way similar to the case of elastic
Maxwell molecules \cite{GS03,SGBD94,SG95}, we find that, for a given
value of the coefficient of restitution, those moments diverge in
time for shear rates larger than a certain critical value
$a_c^*(\alpha)$, which decreases as $\alpha$ decreases. Below this
critical value, the (scaled) fourth-degree moments have well-defined
values in the long time limit. The paper is closed in Section
\ref{sec5} with some concluding remarks.

\section{The Boltzmann equation for  IMM. Collisional moments\label{sec2}}
In the absence of external forces, the inelastic Boltzmann equation
for  IMM reads
\beq
\left(\partial_t
+\mathbf{v}\cdot\nabla\right)f(\mathbf{r},\mathbf{v};t)=J[\mathbf{v}|f,f],
\label{2.1}
\eeq
where the Boltzmann collision operator $J[\mathbf{v}|f,f]$ is given
by \cite{NK03}
\begin{equation}
J\left[{\bf v}_{1}|f,f\right] =\frac{\omeganu}{n\Omega_d} \int
\dd{\bf v}_{2}\int \dd\widehat{\boldsymbol{\sigma}} \left[
\alpha^{-1}f({\bf v}_{1}')f({\bf v}_{2}')-f({\bf v}_{1})f({\bf
v}_{2})\right] .
\label{1}
\end{equation}
Here,
\beq
n=\int\dd \mathbf{v} f(\mathbf{v})
\eeq
is the number density, $\omeganu$ is the collision frequency
(assumed to be independent of $\alpha$),
$\Omega_d=2\pi^{d/2}/\Gamma(d/2)$ is the total solid angle in $d$
dimensions, and $\alpha\leq 1$ refers to the constant coefficient of
restitution. In addition, the primes on the velocities denote the
initial values $\{{\bf v}_{1}^{\prime}, {\bf v}_{2}^{\prime}\}$ that
lead to $\{{\bf v}_{1},{\bf v}_{2}\}$ following a binary collision:
\begin{equation}
\label{3}
{\bf v}_{1}^{\prime}={\bf v}_{1}-\frac{1}{2}\left( 1+\alpha
^{-1}\right)(\widehat{\boldsymbol{\sigma}}\cdot {\bf
g})\widehat{\boldsymbol {\sigma}}, \quad {\bf v}_{2}^{\prime}={\bf
v}_{2}+\frac{1}{2}\left( 1+\alpha^{-1}\right)
(\widehat{\boldsymbol{\sigma}}\cdot {\bf
g})\widehat{\boldsymbol{\sigma}}\;,
\end{equation}
where ${\bf g}={\bf v}_1-{\bf v}_2$ is the relative velocity of the
colliding pair and $\widehat{\boldsymbol{\sigma}}$ is a unit vector
directed along the centers of the two colliding particles. Apart
from $n$, the basic moments of $f$ are the flow velocity
\beq
\mathbf{u}=\frac{1}{n}\int\dd \mathbf{v} \mathbf{v}f(\mathbf{v})
\eeq
and the granular temperature
\beq
T=\frac{m}{dn}\int\dd \mathbf{v}\, V^2 f(\mathbf{v}),
\label{granT}
\eeq
where $\mathbf{V}=\mathbf{v}-\mathbf{u}(\mathbf{r})$ is the peculiar
velocity. The momentum and energy fluxes are characterized by the
pressure tensor
\beq
P_{ij}=m\int\dd \mathbf{v} \,V_i V_j f(\mathbf{v})
\label{Pij}
\eeq
and the heat flux
\beq
\mathbf{q}=\frac{m}{2}\int\dd \mathbf{v}\,
V^2\mathbf{V}f(\mathbf{v}).
\label{qi}
\eeq
Finally, the rate of energy dissipated due to collisions defines the
cooling rate $\zeta$ as
\beq
\zeta=-\frac{m}{dnT}\int\dd \mathbf{v}\, V^2 J[\mathbf{v}|f,f].
\label{zeta}
\eeq

 The main advantage of the Boltzmann equation for Maxwell
models (both elastic and inelastic) is that the (collisional)
moments of $J$ can be exactly evaluated in terms of the moments of
$f$, without the explicit knowledge of the latter \cite{TM80}. This
property has been recently exploited \cite{GS07} to obtain the
detailed expressions for all the third- and fourth-degree
collisional moments as functions of $\alpha$ in $d$ dimensions. In
order to get the collisional moments, it is convenient to introduce
the Ikenberry polynomials \cite{TM80} $Y_{2r|i_1i_2\ldots
i_s}(\mathbf{V})$ of degree $2r+s$. The Ikenberry polynomials  of
degree smaller than or equal to four are
\beq
Y_{0|0}(\mathbf{V})=1,\quad Y_{0|i}(\mathbf{V})=V_i,
\label{X0}
\eeq
\beq
Y_{2|0}(\mathbf{V})=V^2,\quad
Y_{0|ij}(\mathbf{V})=V_iV_j-\frac{1}{d}V^2\delta_{ij},
\label{X1}
\eeq
\beq
Y_{2|i}(\mathbf{V})=V^2 V_i,\quad
Y_{0|ijk}(\mathbf{V})=V_iV_jV_k-\frac{1}{d+2}V^2\left(V_i\delta_{jk}+V_j\delta_{ik}+V_k\delta_{ij}\right),
\label{X2}
\eeq
\beq
Y_{4|0}(\mathbf{V})=V^4,\quad
Y_{2|ij}(\mathbf{V})=V^2\left(V_iV_j-\frac{1}{d}V^2\delta_{ij}\right),
\label{X3}
\eeq
\beqa
Y_{0|ijk\ell}(\mathbf{V})&=&V_iV_jV_kV_\ell-\frac{1}{d+4}V^2
\left(V_iV_j\delta_{k\ell}+V_iV_k\delta_{j\ell}+V_iV_\ell\delta_{jk}
+V_jV_k\delta_{i\ell}+V_jV_\ell\delta_{ik}+V_kV_\ell\delta_{ij}\right)\nn
&&+\frac{1}{(d+2)(d+4)}V^4\left(\delta_{ij}\delta_{k\ell}+\delta_{ik}\delta_{j\ell}+\delta_{i\ell}\delta_{jk}\right)\nn
&=&V_iV_jV_kV_\ell-\frac{1}{d+4}\left[Y_{2|ij}(\mathbf{V})\delta_{k\ell}+Y_{2|ik}(\mathbf{V})\delta_{j\ell}
+Y_{2|i\ell}(\mathbf{V})\delta_{jk}
+Y_{2|jk}(\mathbf{V})\delta_{i\ell}\right.\nn &&\left.
+Y_{2|j\ell}(\mathbf{V})\delta_{ik}
+Y_{2|k\ell}(\mathbf{V})\delta_{ij}\right]-\frac{1}{d(d+2)}V^4\left(\delta_{ij}\delta_{k\ell}+\delta_{ik}\delta_{j\ell}+\delta_{i\ell}\delta_{jk}\right).
\label{X4}
\eeqa
The corresponding velocity moments $M_{2r|i_1i_2\ldots i_s}$ and
collisional moments $J_{2r|i_1i_2\ldots i_s}$ are defined,
respectively, as
\beq
M_{2r|i_1i_2\ldots i_s}=\int\dd\mathbf{v}\, Y_{2r|i_1i_2\ldots
i_s}(\mathbf{V})f(\mathbf{v}),
\label{M2rs}
\eeq
\beq
J_{2r|i_1i_2\ldots i_s}=\int\dd\mathbf{v}\, Y_{2r|i_1i_2\ldots
i_s}(\mathbf{V})J[\mathbf{v}|f,f].
\label{J2rs}
\eeq
In particular, $M_{0|0}=n$,  $M_{0|i}=0$ (by definition of the
peculiar velocity), $M_{2|0}=pd/m$, where $p=nT$ is the hydrostatic
pressure, $M_{0|ij}=(P_{ij}-p\delta_{ij})/m$,  and $M_{2|i}=2q_i/m$.
Moreover, conservation of mass and momentum implies $J_{0|0}=0$ and
$J_{0|i}=0$, respectively, while $J_{2|0}=-\zeta M_{2|0}$.

The explicit expressions for the collisional moments
$J_{2r|i_1i_2\ldots i_s}$ for $2r+s\leq 4$ are \cite{GS07}
\beq
J_{2|0}=-\nutz M_{2|0},\quad J_{0|ij}=-\nuzt M_{0|ij},
\label{X5}
\eeq
\beq
J_{2|i}=-\nuto M_{2|i}, \quad J_{0|ijk}=-\nuzth M_{0|ijk},
\label{X7}
\eeq
\beq
J_{4|0}=-\nufz M_{4|0}+\lambda_1 n^{-1}M_{2|0}^2-\lambda_2
n^{-1}M_{0|ij}M_{0|ji},
\label{X9}
\eeq
\beq
J_{2|ij}=-\nutt M_{2|ij}+\lambda_3 n^{-1}M_{2|0}M_{0|ij}-\lambda_4
n^{-1}\left(M_{0|ik}M_{0|kj}-\frac{1}{d}M_{0|k\ell}M_{0|\ell
k}\delta_{ij}\right),
\label{X11}
\eeq
\beqa
J_{0|ijk\ell}&=&-\nu_{0|4}M_{0|ijk\ell}+\lambda_5
n^{-1}\left[M_{0|ij}M_{0|k\ell}+
M_{0|ik}M_{0|j\ell}+M_{0|i\ell}M_{0|jk}-\frac{2}{d+4}\left(
M_{0|ip}M_{0|pj}\delta_{k\ell}\right.\right. \nn
&&\left.+M_{0|ip}M_{0|pk} \delta_{j\ell}
+M_{0|ip}M_{0|p\ell}\delta_{jk}+M_{0|jp}M_{0|pk}\delta_{i\ell}
+M_{0|jp}M_{0|p\ell}\delta_{ik}+M_{0|kp}M_{0|p\ell}\delta_{ij}\right)\nn
&&\left.
+\frac{2}{(d+2)(d+4)}M_{0|pq}M_{0|qp}\left(\delta_{ij}\delta_{k\ell}+
\delta_{ik}\delta_{j\ell}+\delta_{i\ell}\delta_{jk}\right) \right].
\label{X12}
\eeqa
In Eqs.\ (\ref{X9})--(\ref{X12}),  the usual summation convention
over repeated indices is assumed. The cooling rate $\zeta$ and the
effective collision frequencies $\nu_{2r|{s}}$  are given by  the
expressions
\beq
\nutz=\frac{d+2}{4d}\left(1-\al^2\right)\nu_0,
\label{X6a}
\eeq
\beq
\nuzt=\nutz+\frac{(1+\alpha)^2}{4}\nu_0,
\label{X6}
\eeq
\beq
\nuto=\frac{3}{2}\nutz+\frac{(1+\al)^2(d-1)}{4d}\nu_0,
\label{X7b}
\eeq
\beq
 \nuzth=\frac{3}{2}\nuzt,
\label{X8}
\eeq
\beq
\nufz=2\nutz+\frac{(1+\al)^2\left(4d-7+6\alpha-3\alpha^2\right)}{16d}\nu_0,
\label{X10}
\eeq
\beq
\nutt=
2\nutz+\frac{(1+\al)^2\left[3d^2+7d-14+3\al(d+4)-6\al^2\right]}{8d(d+4)}\nu_0,
\label{X15}
\eeq
\beq
\nuzf=2\nutz+\frac{(1+\al)^2\left[d^3+9d^2+17d-9+3\al(d+4)-3\al^2\right]}{2d(d+4)(d+6)}\nu_0,
\label{X18}
\eeq
where we have called $\nu_0\equiv 2\nu/(d+2)$. According to Eqs.\
\eqref{X5} and \eqref{X6}, $\nu_0$ represents the effective
collision frequency associated with the shear viscosity in the
elastic limit \cite{CC70}.
 Finally, the cross coefficients $\lambda_i$ in Eqs.\
\eqref{X9}--\eqref{X12} are
\beq
\lambda_1=\frac{(1+\al)^2(d+2)\left(4d-1-6\al+3\al^2\right)}{16d^2}\nu_0,
\label{X14}
\eeq
\beq
\lambda_2=\frac{(1+\al)^2\left(1+6\al-3\al^2\right)}{8d}\nu_0,
\label{X13}
\eeq
\beq
\lambda_3=\frac{(1+\al)^2\left[d^2+5d-2-3\al(d+4)+6\al^2\right]}{8d^2}\nu_0,
\label{X17}
\eeq
\beq
\lambda_4=\frac{(1+\al)^2\left[2-d+3\al(d+4)-6\al^2\right]}{4d(d+4)}\nu_0,
\label{X16}
\eeq
\beq
\lambda_5=\frac{(1+\al)^2\left[d^2+7d+9-3\al(d+4)+3\al^2\right]}{2d(d+4)(d+6)}\nu_0.
\label{X19}
\eeq

The above results hold independently of the specific form of the
collision frequency $\nu_0$. On physical grounds, $\nu_0\propto n$.
In the case of {\emph{elastic}} Maxwell molecules, $\nu_0$ is
independent of temperature. The extension of this feature to the
inelastic case defines Model A. On the other hand, one can assume
that $\nu_0$ is an increasing function of temperature (Model B). In
particular $\nu_0(T)\propto n T^{1/2}$ makes Model B mimic the
properties of IHS.

\section{Uniform shear flow. Rheological properties}
\label{sec3}

\subsection{Hierarchy of moment equations}
Let us assume that the gas is under the USF. As said in the
Introduction, this state is macroscopically defined by a constant
density $n$, a spatially uniform temperature $T(t)$, and a linear
flow velocity $\mathbf{u}(y)=a y\widehat\mathbf{x}$ \cite{GS03}. At
a microscopic level, the USF is characterized by a velocity
distribution function that becomes \emph{uniform} in the local
Lagrangian frame, i.e.,
\beq
f(\mathbf{r},\mathbf{v};t)=f(\mathbf{V},t).
\eeq
In this frame, the Boltzmann equation \eqref{2.1} reduces to
\beq
\partial_t f(\mathbf{V})-aV_y\frac{\partial}{\partial
V_x}f(\mathbf{V})=J[\mathbf{V}|f,f].
\label{W1}
\eeq
Equation \eqref{W1} is invariant under the transformations
\beq
(V_x,V_y)\to (-V_x,-V_y),
\label{VxVy}
\eeq
\beq
V_j\to -V_j,\quad j\neq x,y.
\label{Vj}
\eeq
This implies that if the initial state $f(\mathbf{V},0)$ is
consistent with the symmetry properties \eqref{VxVy} and \eqref{Vj}
so is the solution to Eq.\ \eqref{W1} at any time $t>0$. Even if one
starts from an initial condition inconsistent with \eqref{VxVy} and
\eqref{Vj}, it is expected that the solution asymptotically tends
for long times to a function compatible with \eqref{VxVy} and
\eqref{Vj}. The investigation of this expectation, at the level of
moments of degree less than or equal to four, is one of the
objectives of this paper.

The properties of uniform temperature and constant density and shear
rate are enforced in computer simulations by applying the
Lees--Edwards boundary conditions \cite{GS03,LE72}, regardless of
the particular interaction model considered. In the case of boundary
conditions representing realistic plates in relative motion, the
corresponding nonequilibrium state is the so-called Couette flow,
where density, temperature, and shear rate are no longer uniform
\cite{TTMGSD01}.

Multiplying both sides of Eq.\ \eqref{W1} by $Y_{r|i_1i_2\ldots
i_s}(\mathbf{V})$ and integrating over $\mathbf{V}$ one gets
\beq
\partial_t M_{r|i_1i_2\ldots i_s}+a N_{r|i_1i_2\ldots i_s}=J_{r|i_1i_2\ldots i_s},
\label{W2}
\eeq
where we have called
\beq
N_{r|i_1i_2\ldots i_s}\equiv \int
\dd\mathbf{V}\,f(\mathbf{V})V_y\frac{\partial}{\partial
V_x}Y_{r|i_1i_2\ldots i_s}(\mathbf{V}).
\label{W3}
\eeq
In particular,
\beq
N_{2|0}=2M_{0|xy}, \quad N_{0|yy}=-\frac{2}{d}M_{0|xy},\quad
N_{0|xy}=M_{0|yy}+\frac{1}{d}M_{2|0},
\label{W4}
\eeq
\beq
N_{0|ij}=M_{0|iy}\delta_{jx}+M_{0|jy}\delta_{ix}+\frac{1}{d}M_{2|0}(\delta_{ix}\delta_{jy}+\delta_{jx}\delta_{iy})
-\frac{2}{d}M_{0|xy}\delta_{ij},
\label{W5}
\eeq
\beq
N_{4|0}=4M_{2|xy}.
\label{W5.2}
\eeq
More in general, since $V_y\partial_{V_x}Y_{r|i_1i_2\ldots
i_s}(\mathbf{V})$ is a polynomial of degree $2r+s$, the quantity
$N_{r|i_1i_2\ldots i_s}$ can be expressed as a linear combination of
moments of the same degree. In addition, thanks to the structure of
the collision operator for IMM, the collisional moments
$J_{r|i_1i_2\ldots i_s}$ only involve moments of degree equal to or
smaller than $2r+s$. Consequently, the hierarchy \eqref{W2} can be
exactly solved in a recursive way. We will call \emph{asymmetric}
moments those that vanish for velocity distributions compatible with
the invariance properties \eqref{VxVy} and \eqref{Vj}. The remaining
moments will be referred to as \emph{symmetric} moments. In
particular, all the moments of odd degree are asymmetric. Among the
moments of even degree, $M_{2r|xz}$ and $M_{2r|xxxy}$, for instance,
are also asymmetric.

In the particular case of the moment $M_{2|0}=nTd/m$, Eq.\
\eqref{W2} becomes
\begin{equation}
\partial_t M_{2|0}+2a M_{0|xy}=-\nutz M_{2|0},
\label{W6}
\end{equation}
where use has been made of Eq.\ \eqref{X5}. This is not but the
balance equation for the energy in the USF. It is convenient to
introduce the \emph{scaled} moments
\begin{equation}
M_{2r|i_1i_2\ldots i_s}^*=\frac{1}{n v_0^{2r+s}}M_{2r|i_1i_2\ldots
i_s}, \quad v_0\equiv
\sqrt{\frac{2T}{m}}=\sqrt{\frac{2M_{2|0}}{dn}},
\label{X1bis}
\eeq
$v_0$
being the thermal speed, and the reduced shear rate
\beq
a^*\equiv \frac{a}{\nu_0}.
\eeq
 In terms of these scaled variables, Eq.\
\eqref{W2} can be rewritten as
\beq
\frac{1}{\nu_0}\partial_t M_{r|i_1i_2\ldots
i_s}^*+a^* N_{r|i_1i_2\ldots i_s}^*-\left(r+\frac{s}{2}\right)\left(\zeta^*+\frac{4}{d}a^*
M_{0|xy}^*\right)M_{r|i_1i_2\ldots i_s}^*=J_{r|i_1i_2\ldots i_s}^*, \label{W2bis}
\eeq
where $\zeta^*\equiv \zeta/\nu_0$ and
\beq
N_{r|i_1i_2\ldots i_s}^*\equiv \frac{1}{n v_0^{2r+s}}N_{r|i_1i_2\ldots i_s},\quad J_{r|i_1i_2\ldots i_s}^*\equiv
\frac{1}{\nu_0n v_0^{2r+s}}J_{r|i_1i_2\ldots i_s}.
\eeq
It is apparent that the evolution equation \eqref{W2bis}
involves the second-degree moment $M_{0|xy}^*=P_{xy}/2p$, which is the (reduced) shear stress. This quantity,
along with the normal stress differences $M_{0|xx}^*=(P_{xx}-p)/2p$ and $M_{0|yy}^*=(P_{yy}-p)/2p$, are the most
relevant ones from a rheological point of view. They will be analyzed in the next subsection.

\subsection{Second-degree moments}

{}From Eq.\ \eqref{W2bis} one gets a coupled set of equations for
the moments $M_{0|xy}^*$ and $M_{0|yy}^*$:
\beq
\frac{1}{\nu_0}\partial_t
M_{0|xy}^*+a^*\left(M_{0|yy}^*+\frac{1}{2}\right)+\left(\omegazt-\frac{4}{d}a^*M_{0|xy}^*\right)M_{0|xy}^*=0,
\label{W7}
\eeq
\beq
\frac{1}{\nu_0}\partial_t
M_{0|yy}^*-\frac{2}{d}a^*M_{0|xy}^*+\left(\omegazt-\frac{4}{d}a^*M_{0|xy}^*\right)M_{0|yy}^*=0,
\label{W8}
\eeq
where we have introduced the (reduced) shifted quantities
\beq
\omega_{2r|s}\equiv \frac{\nu_{2r|s}-(r+s/2)\nutz}{\nu_0},
\eeq
so that
\beq
\omegazt=\frac{(1+\alpha)^2}{4} .
\eeq
To close the set, we need in
general the evolution equation for the reduced shear rate $a^*$.
{}From Eq.\ \eqref{W6} is straightforward to obtain
\beq
\frac{1}{\nu_0}\partial_t a^*=a^*\left(\zeta^*+\frac{4}{d}a^*
M_{0|xy}^*\right)\frac{\partial\ln \nu_0}{\partial\ln T}.
\label{W8.1}
\eeq

\subsubsection{Model A. Hydrodynamic solution}

In Model A the collision frequency $\nu_0$ is independent of
temperature and thus it is a constant. Consequently, $\partial_t
a^*=0$ so that the reduced shear rate $a^*$ remains in its initial
value (regardless of the value of the coefficient of restitution
$\alpha$) and represents a control parameter measuring the departure
of the system from the homogeneous cooling state.

As in the elastic case \cite{GS03,TM80}, it is easy to check that,
after a certain kinetic regime lasting a few collision times, the
scaled moments $M_{0|xy}^*$ and $M_{0|yy}^*$ reach well-defined
stationary values, which are nonlinear functions of  $\alpha$ and
 $a^*\equiv a/\nu_0$. {}From Eqs.\ \eqref{W7} and \eqref{W8}, one
 has
\beq
-M_{0|xy}^*\left(1-\frac{4}{d}\widetilde{a}M_{0|xy}^*\right)=\widetilde{a}\left(M_{0|yy}^*+\frac{1}{2}\right),
\label{W10}
\eeq
\beq
M_{0|yy}^*\left(1-\frac{4}{d}\widetilde{a}M_{0|xy}^*\right)=\frac{2}{d}\widetilde{a}M_{0|xy}^*
\label{W11}
\eeq
for the stationary values, where we have defined
\beq
\widetilde{a}\equiv \frac{a^*}{\nubar}=\frac{4a^*}{(1+\al)^2}.
\label{W12}
\eeq
The solution to the set of equations (\ref{W10}) and (\ref{W11}) is
\beq
M_{0|yy}^*=-\frac{\gamma(\widetilde{a})}{1+2\gamma(\widetilde{a})},\quad
M_{0|xy}^*=-\frac{d}{2}\frac{\gamma(\widetilde{a})}{\widetilde{a}}=-\frac{\widetilde{a}/2}{\left[
1+2\gamma(\widetilde{a})\right]^2},
\label{W13}
\eeq
where
\beq
\gamma(\widetilde{a})=\frac{2}{3}\sinh^2\left[\frac{1}{6}\cosh^{-1}\left(1+\frac{27}{d}\widetilde{a}^2\right)\right]
\label{W15}
\eeq
is the real root of the cubic equation
\beq
\gamma(1+2\gamma)^2=\frac{\widetilde{a}^2}{d}.
\label{W14}
\eeq

Note that the reduced second-degree moments depend on $\alpha$ and
$a^*$ through the scaled quantity $\widetilde{a}$ only. {}From Eq.\
\eqref{W2bis} it is also easy to prove that, for long times, the
normal stresses $M_{zz}^*,\ldots,M_{dd}^*$ along directions
orthogonal to the shear plane  $xy$ are   equal to $M_{yy}^*$.
Consequently, $M_{0|xx}^*=-(d-1)M_{0|yy}^*$. Analogously, the
asymmetric second-degree moments (i.e., all the off-diagonal
elements $M_{0|ij}^*$ except $M_{0|xy}^*$) vanish.

It is convenient to define a nonlinear shear viscosity $\eta^*$ and
a viscometric function $\Psi$ as
\beq
\eta^*(a^*)=-\frac{\nu_0}{p}\frac{P_{xy}}{a}=-2\frac{M_{0|xy}^*}{a^*},
\label{W24}
\eeq
\beq
\Psi(a^*)=\frac{\nu_0^2}{p}\frac{P_{xx}-P_{yy}}{a^2}=2\frac{M_{0|xx}^*-M_{0|yy}^*}{{a^*}^2}.
\label{W35}
\eeq
{}From Eqs. (\ref{W13}) and (\ref{W14}), we have
\beq
\eta^*(a^*)=\left(\frac{2}{1+\al}\right)^2\frac{1}{\left[1+2\gamma(\widetilde{a})\right]^2},
\label{W25}
\eeq
\beq
\Psi(a^*)=\left(\frac{2}{1+\al}\right)^4\frac{2}{\left[1+2\gamma(\widetilde{a})\right]^3}.
\label{W36}
\eeq
Interestingly enough, the combination
\beq
\frac{[\eta^*(a^*)]^3}{[\Psi(a^*)]^2}=\left(\frac{1+\al}{4}\right)^2
\eeq
is independent of the shear rate. Moreover, in the limit of small
shear rate (for fixed $\alpha$), Eq.\ \eqref{W14} implies that
$\gamma\to 0$, so that Eqs.\ \eqref{W25} and \eqref{W36} reduce to
\beq
\eta^*(0)=\frac{4}{(1+\al)^2},\quad \Psi(0)=\frac{32}{(1+\al)^4}.
\eeq
The quantities $\eta^*(0)$ and $\Psi(0)$ are the NS shear viscosity
and the Burnett value of the viscometric function, respectively, of
Model A.

It is important to remark that, although the scaled moments reach
stationary values,  the system is not in general in a steady state
since the temperature  changes in time. Actually, inserting the
second expression of (\ref{W13}) into Eq.\ (\ref{W6}), we get
\beq
\frac{1}{\nu_0}\partial_t \ln
T=-2\nubar\left[\gamma_s-\gamma(\widetilde{a})\right],
\label{W17}
\eeq
where we have called
\beq
\gamma_s\equiv\frac{\nutz^*}{2\nubar}=\frac{d+2}{2d}\frac{1-\al}{1+\al}.
\label{W18b}
\eeq
 Equation
(\ref{W17}) shows that $T(t)$ either grows or decays exponentially.
The first situation occurs if $\gamma(\widetilde{a})>\gamma_s$. In
that case, the imposed shear rate is sufficiently large (or the
inelasticity is sufficiently low) as to make the viscous heating
effect dominate over the inelastic cooling effect. The opposite
happens if $\gamma(\widetilde{a})<\gamma_s$. A perfect balance
between both effects takes place if
$\gamma(\widetilde{a})=\gamma_s$. Inserting this condition into Eq.\
(\ref{W14}) one gets the steady-state point
\beq
{{a}_s^*}=\omegazt\sqrt{d\gamma_s}(1+2\gamma_s)=\sqrt{\frac{d+2}{2}(1-\al^2)}\frac{d+1-\al}{2d}.
\label{W16}
\eeq
In this  state, the $\alpha$-dependence of the rheological
properties is
\beq
\eta_s^*=\eta^*(a^*_s)=\left(\frac{d}{d+1-\alpha}\right)^2,\quad
\Psi_s=\Psi(a^*_s)=\frac{4}{1+\al}\left(\frac{d}{d+1-\alpha}\right)^3.
\label{W26}
\eeq
Equations \eqref{W16} and \eqref{W26} agree with the results
reported in Ref.\ \cite{G03}, while the more general expressions
\eqref{W25} and \eqref{W36} had not been previously derived.

\begin{figure}[tbp]
\includegraphics[width=.60 \columnwidth]{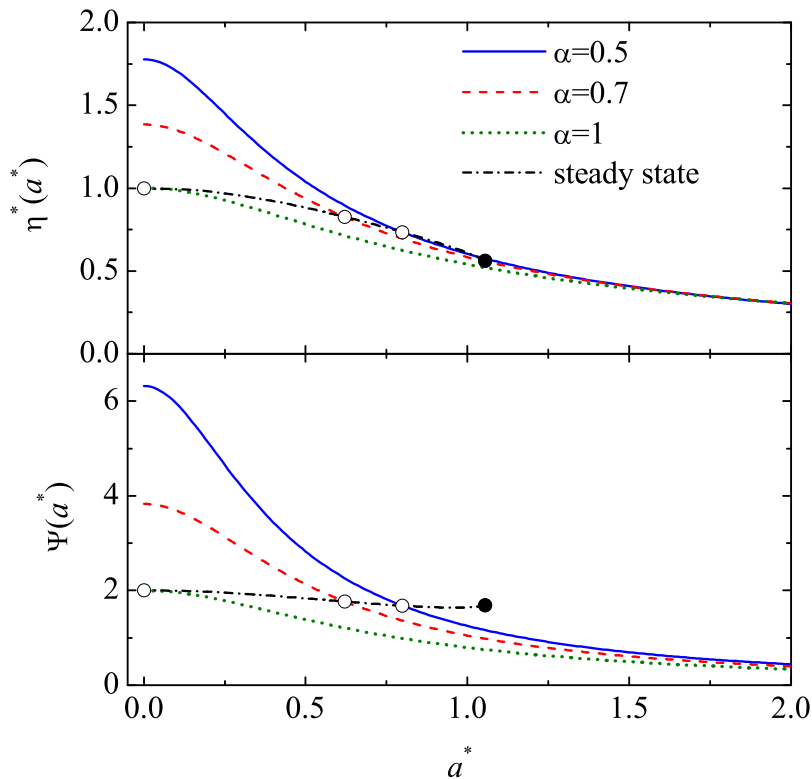}
\caption{(Color online) Plot of $\eta^*(a^* )$ (top panel) and
$\Psi(a^*)$ (bottom panel) as  functions of $a^{\ast }$ for $d=3$
and $\alpha =0.5$ (solid lines), $\alpha =0.7$ (dashed lines), and
$\alpha =1$ (dotted lines). The  dash-dotted lines are the loci of
steady-state points $(a_{s}^{\ast },\eta _{s}^{\ast })$ and
$(a^*_s,\Psi_s)$. They intercept the curves representing $\eta
^{\ast}(a^{\ast })$ and $\Psi(a^*)$ at the steady-state values
indicated by circles. { Note that the loci  end at the points
$(a_{s}^{\ast },\eta _{s}^{\ast })=(1.054,0.563)$ and
$(a^*_s,\Psi_s)=(1.054,1.688)$ corresponding to $\alpha=0$
(represented by filled circles)}.
\label{fig1}}
\end{figure}
\begin{figure}[tbp]
\includegraphics[width=.60 \columnwidth]{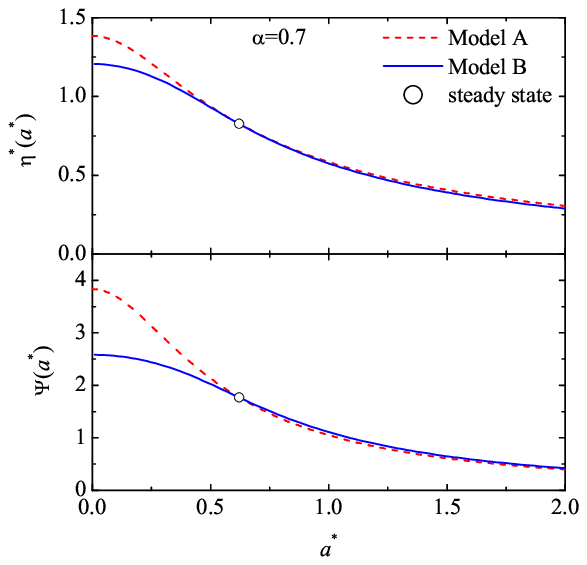}
\caption{(Color online) Plot of $\eta^*(a^* )$ (top panel) and
$\Psi(a^*)$ (bottom panel) as  functions of $a^{\ast }$ for $d=3$
and $\alpha =0.7$. The dashed and solid lines correspond to Model A
[Eqs.\ \protect\eqref{W25} and \protect\eqref{W25}] and Model B with
$\nu_0(T)\propto T^{1/2}$ [numerical solution of Eqs.\
\protect\eqref{W7}--\protect\eqref{W8.1}], respectively. The circles
represent the steady-state points $(a_{s}^{\ast },\eta _{s}^{\ast
})$ and $(a^*_s,\Psi_s)$, which are common to Models A and B.
\label{fig2}}
\end{figure}

Figure \ref{fig1} shows the shear-rate dependence of the rheological
functions $\eta^*(a^*)$ and $\Psi(a^*)$ for $d=3$ and three values
of the coefficient of restitution: $\alpha=0.5$ (highly inelastic
system), $\alpha=0.7$ (moderately inelastic system), and $\alpha=1$
(elastic system). The steady-state points $(a_s^*,\eta_s^*)$ and
$(a_s^*,\Psi_s)$ are also represented by circles for each one of the
values of $\alpha$. Given a value of $\alpha$, the steady-state
point splits each curve into two branches: the one with $a^*>a_s^*$
corresponds to $\gamma(\widetilde{a})>\gamma_s$ and so the
temperature increases in time, while the branch with $a^*<a^*_s$
corresponds to states with a decreasing temperature. We observe
that, for a given value of $\alpha$, both $\eta^*(a^*)$ and
$\Psi(a^*)$ decrease with increasing shear rate. In the region of
high shear rates (say $a^*>1.5$), the curves are practically
insensitive to the value of the coefficient of restitution. For
small and moderate shear rates, however, the influence of $\alpha$
is noticeable: at a given value of the reduced shear rate $a^*$, the
rheological quantities $\eta^*(a^*)$ and $\Psi(a^*)$ increase as
$\alpha$ decreases. On the other hand, the steady-state values
$\eta_s^*$ and $\Psi_s^*$ decrease with increasing dissipation.

\subsubsection{Model B. Steady-state
solution}

In Model B the collision frequency $\nu_0(T)$ is an increasing
function of temperature, and so the reduced shear rate $a^*$ is not
constant. The corresponding steady-state solution is obtained from
Eqs.\ (\ref{W6})--(\ref{W8.1}) by setting $\partial_t\to 0$. It is
given again by Eqs.\ (\ref{W13})--(\ref{W14}), except that now
$\gamma(\widetilde{a})\to \gamma_s$ and $\widetilde{a}\to
a_s^*/\omegazt$, where $\gamma_s$ and $a_s^*$ are given by Eqs.\
\eqref{W18b} and \eqref{W16}, respectively. Therefore, the
steady-state results are ``universal'' in the sense that they hold
both for Model A and Model B, regardless of the precise dependence
$\nu_0(T)$.

In order to have $M_{0|xy}^*(a^*)$ and $M_{0|yy}^*(a^*)$ in Model B,
one has to solve numerically the nonlinear coupled set
\eqref{W7}--\eqref{W8.1},  discard the kinetic stage of the
evolution, and eliminate time in favor of $a^*$ \cite{SGD04}. The
resulting rheological curves are illustrated in Fig.\ \ref{fig2} at
$\alpha=0.7$ and for the choice $\nu_0(t)\propto T^{1/2}$ in Model
B. Comparison with the analytical results corresponding to Model A
shows that the influence of the temperature dependence of $\nu_0$ on
the rheological properties is only significant for reduced shear
rates smaller than the steady-state one. Since in this paper we want
to focus on analytical results, henceforth we will only consider
Model A, except for what concerns the common steady state of Models
A and B.

\section{Third- and fourth-degree moments}
\label{sec4}
 In this Section we will  analyze, in the context of Model A, the time evolution and the stationary values  of the (scaled) third- and
fourth-degree moments in the USF problem. The results will depend on
both the reduced shear rate $a^*$ and the coefficient of restitution
$\alpha$, while they only depend on the latter in the common steady
state.

Let us assume that the scaled second-degree moments have reached
their stationary values given by Eq.\ \eqref{W13}. Therefore, Eq.\
\eqref{W2bis} becomes
\beq
\partial_s M_{r|i_1i_2\ldots i_s}^*+a^*
N_{r|i_1i_2\ldots
i_s}^*-\left(r+\frac{s}{2}\right)\left[\zeta^*-2\omegazt\gamma(\widetilde{a})\right]M_{r|i_1i_2\ldots
i_s}^*=J_{r|i_1i_2\ldots i_s}^*,
\label{W19}
\eeq
where $\dd s=\nu_0\dd t$. In what follows we will particularize to a
three-dimensional gas ($d=3$).

\subsection{Third-degree moments}
As said in the preceding Section, all the third-degree moments are
asymmetric and so they are expected to vanish for long times. Here
we want to confirm this expectation and get the corresponding
relaxation rates.

In a three-dimensional system, there are 10  independent
third-degree moments. Here we take
\begin{equation}
\{M_{2|x}^*,
M_{2|y}^*,M_{2|z}^*,M_{0|xxy}^*,M_{0|xxz}^*,M_{0|xyy}^*,M_{0|yyz}^*,M_{0|xzz}^*,M_{0|yzz}^*,
M_{0|xyz}^*\} .
\label{3.4:M}
\end{equation}
{}From Eq.\ (\ref{W19}) and making use of the third-degree
collisional moments \eqref{X7} and of the definition \eqref{W3}, one
gets the following set of equations:
\begin{equation}
\left[\partial_{s}+\omegazth+3\omegazt\gamma(\widetilde{a})\right]\left(\frac{1}{4}M_{0|xxz}^*+M_{0|yyz}^*
\right)=0,
\label{3.4:c7}
\end{equation}
\begin{equation}
\left[\partial_{s}+\omegazth+3\omegazt\gamma(\widetilde{a})\right]\left(\frac{1}{4}M_{0|xxy}^*
+M_{0|yzz}^*\right)=0,
\label{3.4:c8}
\end{equation}
\begin{equation}
\left(
\begin{array}{ccc}
\partial_{s}+\omegato+3\omegazt\gamma(\widetilde{a})&0&2a^*\\
0&\partial_{s}+\omegazth+3\omegazt\gamma(\widetilde{a})&-\frac{2}{5}a^*\\
\frac{1}{5}a^*&a^*&\partial_{s}+\omegazth+3\omegazt\gamma(\widetilde{a})
\end{array}\right)
\cdot \left(
\begin{array}{c}
M_{2|z}^*\\
M_{0|yyz}^*\\
M_{0|xyz}^*
\end{array}\right)
=\left(\begin{array}{c}
0\\
0\\
0
\end{array}\right),
\label{3.4:c9}
\end{equation}
\begin{eqnarray}
&&\left(
\begin{array}{cccc}
\partial_{s}+\omegato+3\omegazt\gamma(\widetilde{a})&\frac{7}{5}a^*&2a^*&0\\
\frac{2}{5}a^*&\partial_{s}+\omegato+3\omegazt\gamma(\widetilde{a})&0&2a^*\\
\frac{8}{25}a^*&0&\partial_{s}+\omegazth+3\omegazt\gamma(\widetilde{a})&\frac{8}{5}a^*\\
0&\frac{8}{25}a^*&-\frac{23}{20}a^*&\partial_{s}+\omegazth+3\omegazt\gamma(\widetilde{a})
\end{array}\right)\nonumber \\
&&\cdot \left(
\begin{array}{c}
M_{2|x}^*\\
M_{2|y}^*\\
M_{0|xxy}^*\\
M_{0|xyy}^*
\end{array}\right)
= a^*\left(\frac{1}{4}M_{0|xxy}^*+M_{0|yzz}^*\right) \left(
\begin{array}{c}
0\\0\\0\\
1
\end{array}\right)
 ,
\label{3.4:c10}
\end{eqnarray}
\begin{equation}
\left[\partial_{s}+\omegazth+3\omegazt\gamma(\widetilde{a})\right]M_{0|xzz}^*=a^*\left(\frac{2}{5}M_{2|y}^*+
\frac{2}{5}M_{0|xxy}^* -M_{0|yzz}^*\right).
\label{3.4:c11}
\end{equation}

The characteristic equations associated with Eqs.\
\eqref{3.4:c7}--\eqref{3.4:c11} are
\beq
\ell=\omegazth+3\omegazt\gamma(\widetilde{a})=\frac{3}{4}(1+\al)^2\left[\gamma(\widetilde{a})+\frac{1}{2}\right],
\eeq
\begin{equation}
\left[\omegazth+3\omegazt\gamma(\widetilde{a})-\ell\right]^2\left[\omegato+3\omegazt\gamma(\widetilde{a})-\ell\right]=\frac{2(\omegazth-\omegato)}{5}{a^*}^2,
\label{3.4:c12}
\end{equation}
\begin{equation}
\left[\omegazth+3\omegazt\gamma(\widetilde{a})-\ell\right]^2\left[\omegato+3\omegazt\gamma(\widetilde{a})-\ell\right]^2=\frac{2(\omegazth-\omegato)}{25}
{a^*}^2
\left[7\omegazth+23\omegato+30\left(3\omegazt\gamma(\widetilde{a})-\ell\right)\right],
\label{3.4:c13}
\end{equation}
where $\ell$ denotes the corresponding eigenvalues. The time
evolution for long times is governed by the eigenvalue
$\ell_{\text{min}}$ with the smallest real part. It can be checked
that $\ell_{\text{min}}$ [which is the smallest real root of the
quartic equation \eqref{3.4:c13}] is positive definite for all
$\alpha$ and $a^*$. Consequently, all the scaled third-degree
moments vanish in the long time limit, as expected by symmetry
arguments. In addition, at a given value of $a^*$, the larger the
inelasticity the longer the relaxation time (which is proportional
to $\ell_{\text{min}}^{-1}$). The shear-rate dependence of
$\ell_{\text{min}}$ is plotted in Fig.\ \ref{fig3} for the same
values of $\alpha$ as considered before. It is interesting to remark
that $\ell_{\text{min}}$ is not a monotonic function of $a^*$,
reaching a minimum value at a certain shear rate.

\begin{figure}[tbp]
\includegraphics[width=.60 \columnwidth]{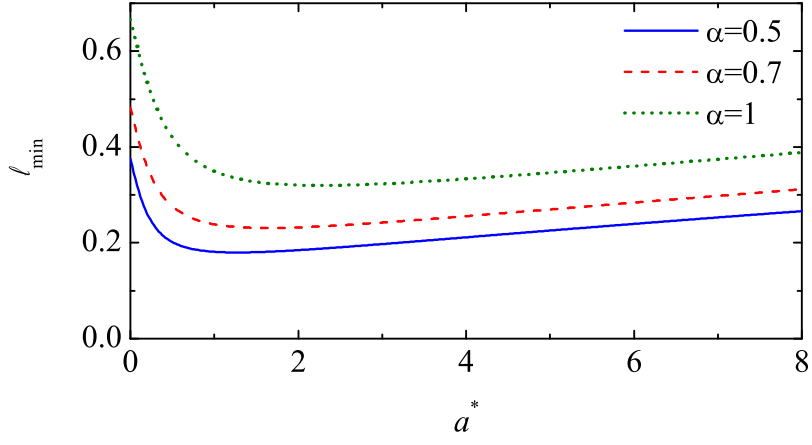}
\caption{(Color online) Plot of the smallest eigenvalue,
$\ell_{\text{min}}$, associated with the time evolution of the
third-degree moments
 as a
function of $a^{\ast }$ for $d=3$ and $\alpha =0.5$ (solid line),
$\alpha =0.7$ (dashed line), and $\alpha =1$ (dotted line).
\label{fig3}}
\end{figure}

\subsection{Fourth-degree moments}
In a three-dimensional system, there are 15 independent
fourth-degree moments, of which 9  are symmetric and 6 are
symmetric, in the sense described at the beginning of Sec.\
\ref{sec3}.

\subsubsection{Asymmetric moments}

\begin{figure}[tbp]
\includegraphics[width=.60 \columnwidth]{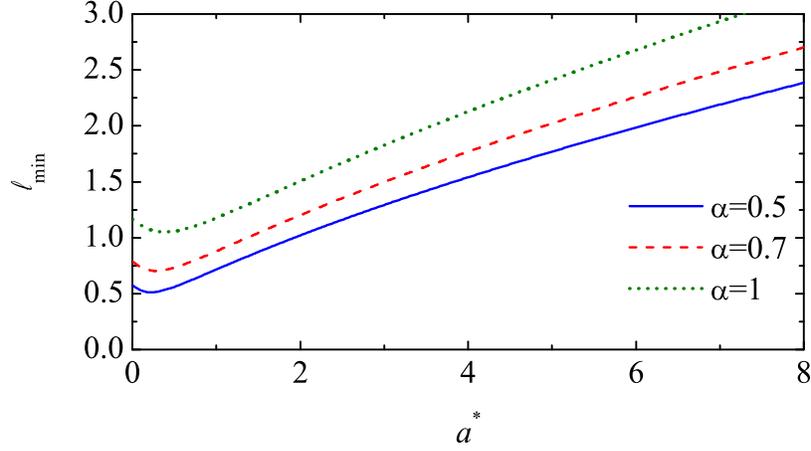}
\caption{(Color online) Plot of the smallest eigenvalue,
$\ell_{\text{min}}$, associated with the time evolution of the
asymmetric fourth-degree moments
 as a
function of $a^{\ast }$ for $d=3$ and $\alpha =0.5$ (solid line),
$\alpha =0.7$ (dashed line), and $\alpha =1$ (dotted line).
\label{fig4}}
\end{figure}

Because of the symmetries of Eq.\ (\ref{W1}), the
 symmetric and  asymmetric moments are uncoupled. Although the
relevant moments are the symmetric ones,  we first analyze the time
evolution of the asymmetric moments, for the sake of completeness.
As the set of  asymmetric moments, we choose
\beq
\{M_{2|xz}^*,M_{2|yx}^*,
M_{0|xxxzy}^*,M_{0|yyyz}^*,M_{0|xzzz}^*,M_{0|yzzz}^*\}.
\label{set4as}
\eeq
They verify the following set of equations:
\begin{equation}
\left[\partial_s+\omegazf+4\nubar\gamma(\widetilde{a})\right](M_{0|yyyz}^*-M_{0|yzzz}^*)=0,
\label{3.4:d9}
\end{equation}
\begin{eqnarray}
&&\left(
\begin{array}{cccc}
\partial_s+\omegatt+4\nubar\gamma(\widetilde{a})&\frac{9}{7}a^*&0&-2a^*\\
\frac{2}{7}a^*&\partial_s+\omegatt+4\nubar\gamma(\widetilde{a})&-2a^*&0\\
0&-\frac{12}{49}a^*&\partial_s+\omegazf+4\nubar\gamma(\widetilde{a})&-\frac{11}{14}a^*\\
-\frac{12}{49}a^*&0&\frac{12}{7}a^*&\partial_s+\omegazf+4\nubar\gamma(\widetilde{a})
\end{array}
\right)
\nonumber\\
&&\cdot \left(
\begin{array}{c}
M_{2|xz}^*\\
M_{2|yz}^*\\
M_{0|xxxz}^*+M_{0|xzzz}^*\\
M_{0|yyyz}^*+M_{0|yzzz}^*
\end{array}
\right) = \frac{1}{2}a^*(M_{0|yyyz}^*-M_{0|yzzz}^*) \left(
\begin{array}{c}
0\\0\\1\\0
\end{array}
\right),
\label{3.4:d10}
\end{eqnarray}
\beq
\left[\partial_s+\omegazf+4\nubar\gamma(\widetilde{a})\right](M_{0|xxxz}^*-M_{0|xzzz}^*)=\frac{7}{2}a^*(M_{0|yyyz}^*+M_{0|yzzz}^*)
-\frac{1}{2}a^*(M_{0|yyyz}^*-M_{0|yzzz}^*).
\label{3.4:d11}
\eeq
The eigenvalues associated with the  time behavior of the asymmetric
fourth-degree moments are
\beq
\ell=\omegazf+4\omegazt\gamma(\widetilde{a})=(1+\al)^2\left[\gamma(\widetilde{a})+\frac{50+7\al-\al^2}{126}\right]
\eeq
and the roots of the characteristic  quartic equation
\begin{equation}
\left[\omegazf+4\omegazt\gamma(\widetilde{a})-\ell\right]^2\left[\omegatt+4\omegazt\gamma(\widetilde{a})-\ell\right]^2=\frac{6(\omegazf-\omegatt)}{49}
{a^*}^2
\left[3\omegazf+11\omegatt+14\left(4\omegazf\gamma(\widetilde{a})-\ell\right)\right].
\label{3.4:c13bis}
\end{equation}
All the eigenvalues have a positive real part. Therefore, all the
asymmetric moments  defined in Eq.\ \eqref{set4as} decay to zero in
the long time limit, the final stage being characterized by the
smallest real root $\ell_{\text{min}}$ of Eq.\ \eqref{3.4:c13bis}.
This eigenvalue is plotted in Fig.\ \ref{fig4} for the same cases as
in Fig.\ \ref{fig3}. Again, a non-monotonic behavior is observed. On
the other hand, for given values of $a^*$ and $\alpha$, the value of
$\ellmin$ corresponding to the third-degree moments is smaller than
the one corresponding to the asymmetric fourth-degree moments.

\subsubsection{Symmetric moments}
 In
parallel to the elastic case \cite{GS03}, we choose the following
set of 9 symmetric moments
\beq
\{M_{4|0}^*,M_{2|xx}^*,
M_{2|yy}^*,M_{2|xy}^*,M_{0|xxxx}^*,M_{0|yyyy}^*,M_{0|zzzz}^*,
M_{0|xxxy}^*,M_{0|xyyy}^*\}.
\label{3.4:Mbis}
\eeq
The combination
\begin{equation}
{\cal M}_9\equiv
3M_{0|xxxx}^*-4M_{0|yyyy}^*-4M_{0|zzzz}^*=\frac{1}{nv_0^4}\int
\dd\mathbf{V} \left(6 V_y^2V_z^2- V_y^4- V_z^4\right)f(\mathbf{V})
\label{3.4:M9}
\end{equation}
is the average of a quantity independent of $V_x$, so the associated
combination $3N_{0|xxxx}^*-4N_{0|yyyy}^*-4N_{0|zzzz}^*$ vanishes.
Moreover, the  combination
$3J_{0|xxxx}^*-4J_{0|yyyy}^*-4J_{0|zzzz}^*=-\nu_{0|4}{\cal M}_9$ due
to the fact that $M_{0|yy}=M_{0|zz}$. Therefore, Eq.\ \eqref{W19}
yields
\beq
\left[\partial_s+\omegazf+4\nubar\gamma(\widetilde{a})\right]\mathcal{M}_9=0.
\label{W22}
\eeq
The solution to this equation is simply
\beq
\mathcal{M}_9(s)=\mathcal{M}_9(0)\ee^{-\ell_9 s},\quad \ell_9 \equiv
\omegazf+4 \nubar\gamma(\widetilde{a}).
\label{W23}
\eeq
Since $\omegazf>0$ \cite{GS07}, one has  $\ell_9>0$ and so
$\mathcal{M}_9$ decays to 0.

The remaining eight moments in (\ref{3.4:Mbis}) are coupled. In
matrix form, Eq.\ \eqref{W19} becomes
\begin{equation}
(\delta_{\sigma \sigma'}\partial_{s}+{\cal L}_{\sigma \sigma'})
{\cal M}_{\sigma'}= {\cal C}_{\sigma}, \quad \sigma =1,\ldots, 8.
\label{3.4:d13}
\end{equation}
Here,  $\bm{\mathcal{M}}$ is a vector made of the following 8
moments
\begin{equation}
\bm{\mathcal{M}}=\left(
\begin{array}{c}
M_{4|0}^*\\M_{2|xx}^*\\M_{2|yy}^*\\M_{0|yyyy}^*\\M_{0|zzzz}^*\\M_{2|xy}^*\\M_{0|xxxy}^*
\\M_{0|xyyy}^*
\end{array}
\right),
\label{3.4:d14}
\end{equation}
and the square matrix  $\bm{\mathcal{L}}$ is
\begin{equation}
\bm{\mathcal{L}}=4\nubar\gamma(\ati) \bm{\mathcal{I}}+
\bm{\mathcal{L}}',
\label{3.4:n1}
\end{equation}
where  $\bm{\mathcal{I}}$ is the $8\times 8$ identity matrix and
\begin{equation}
\bm{\mathcal{L}}'=\left(
\begin{array}{cccccccc}
\omegafz&0&0&0&0&4a^*&0&0\\
0&
\omegatt&0&0&0&\frac{32}{21}a^*&2a^*&0\\
0&0&
\omegatt&0&0&-\frac{10}{21}a^*&0&2a^*\\
0&0&0&
\omegazf&0&-\frac{96}{245}a^*&0&-\frac{12}{7}a^*\\
0&0&0&0&
\omegazf&\frac{24}{245}a^*&\frac{12}{7}a^*&\frac{12}{7}a^*\\
\frac{7}{15}a^*&\frac{2}{7}a^*&\frac{9}{7}a^*&-\frac{7}{3}a^*
&-\frac{1}{3}a^*&
\omegatt&0&0\\
0&\frac{15}{49}a^*&-\frac{6}{49}a^*&-\frac{5}{2}a^*&-\frac{5}{14}a^*&0&
\omegazf&0\\
0&-\frac{6}{49}a^*&\frac{15}{49}a^*&2a^*&\frac{1}{7}a^*&0&0&
\omegazf
\end{array}
\right).
\label{3.4:d15}
\end{equation}
In addition, $\bm{\mathcal{C}}$ is a vector of elements made of
second-degree moments, namely
\beq
\mathcal{C}_1= \frac{9}{4}{\lambda}^*_1-
{\lambda}_2^*\left({M_{0|xx}^*}^2+{M_{0|yy}^*}^2+{M_{0|zz}^*}^2+2{M_{0|xy}^*}^2\right),
\label{W27}
\eeq
\beq
\mathcal{C}_2= \frac{3}{2}{\lambda}^*_3M_{0|xx}^*-\frac{1}{3}
{\lambda}_4^*\left(2{M_{0|xx}^*}^2-{M_{0|yy}^*}^2-{M_{0|zz}^*}^2+{M_{0|xy}^*}^2\right),
\label{W28}
\eeq
\beq
\mathcal{C}_3= \frac{3}{2}{\lambda}^*_3M_{0|yy}^*-\frac{1}{3}
{\lambda}_4^*\left(2{M_{0|yy}^*}^2-{M_{0|xx}^*}^2-{M_{0|zz}^*}^2+{M_{0|xy}^*}^2\right),
\label{W29}
\eeq
\beq
\mathcal{C}_4=\frac{1}{35}
{\lambda}^*_5\left(51{M_{0|yy}^*}^2+6{M_{0|xx}^*}^2+6{M_{0|zz}^*}^2-48{M_{0|xy}^*}^2\right),
\label{W30}
\eeq
\beq
\mathcal{C}_5=\frac{1}{35}
{\lambda}^*_5\left(51{M_{0|zz}^*}^2+6{M_{0|xx}^*}^2+6{M_{0|yy}^*}^2+12{M_{0|xy}^*}^2\right),
\label{W31}
\eeq
\beq
\mathcal{C}_6= \frac{3}{2}{\lambda}^*_3M_{0|xy}^*-
{\lambda}_4^*M_{0|xy}^*\left({M_{0|xx}^*}+{M_{0|yy}^*}\right),
\label{W32}
\eeq
\beq
\mathcal{C}_7=\frac{3}{7}
{\lambda}^*_5{M_{0|xy}^*}\left(5{M_{0|xx}^*}-2{M_{0|yy}^*}\right),
\label{W33}
\eeq
\beq
\mathcal{C}_8=\frac{3}{7}
{\lambda}^*_5{M_{0|xy}^*}\left(5{M_{0|yy}^*}-2{M_{0|xx}^*}\right).
\label{W34}
\eeq

\begin{figure}[tbp]
\includegraphics[width=.60 \columnwidth]{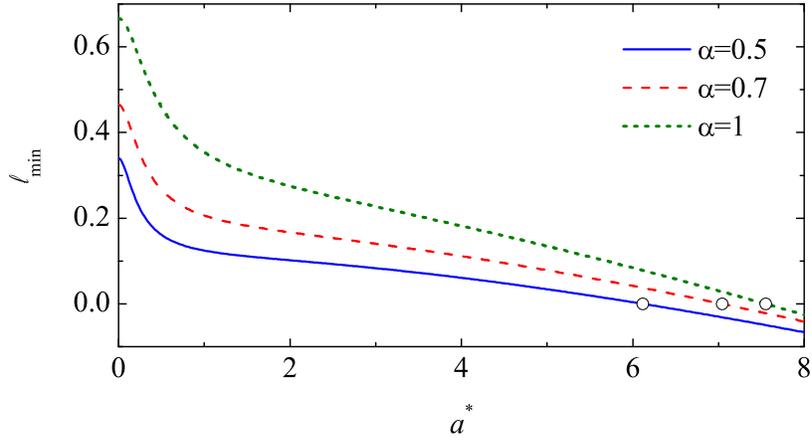}
\caption{(Color online) Plot of the smallest eigenvalue, $\ellmin$,
associated with the time evolution of the symmetric fourth-degree
moments as a function of $a^{\ast }$ for $\alpha =0.5$ (solid line),
$\alpha=0.7$ (dashed line), and $\alpha =1$ (dotted line).  The
circles indicate the location of the corresponding values of the
critical shear rate.
\label{fig5}}
\end{figure}

The solution of Eq.\ (\ref{3.4:d13}) can be written as
\begin{equation}
\bm{\mathcal{M}}(s)=\ee^{-\bm{\mathcal{L}}s}\cdot
[\bm{\mathcal{M}}(0)-\bm{\mathcal{M}}_\infty]+\bm{\mathcal{M}}_\infty
,
\label{3.4:d17}
\end{equation}
where
\begin{equation}
\bm{\mathcal{M}}_\infty=\bm{\mathcal{L}}^{-1}\cdot\bm{\mathcal{C}} .
\label{3.4:d18}
\end{equation}
Similarly to the cases discussed above, the long time behavior of
${\cal M}_\sigma$ ($\sigma=1,\ldots,8$) is governed by the
eigenvalue $\ell_{\text{min}}$ of the matrix $\bm{\mathcal{L}}$ with
the smallest real part. We have checked that $\ell_{\text{min}}$ is
a real quantity that, for a given value of $\alpha$, monotonically
decreases with increasing shear rate. It is plotted in  Fig.\
\ref{fig5} for $\alpha=0.5$, $0.7$, and $1$. The most important
feature of Fig.\ \ref{fig5} is that, for any given value of
$\alpha$, $\ell_{\text{min}}$  becomes negative for shear rates
larger than a certain ``critical'' value $a_c^*(\alpha)$. This means
that, if $a^*>a_c^*$, the symmetric fourth-degree moments
exponentially grow in time. This singular behavior of the scaled
moments implies that the velocity distribution function (scaled with
the thermal speed) develops an algebraic high-velocity tail in the
long time limit. It is interesting to remark that this effect is
also present in the elastic limit, where it has been extensively
studied \cite{GS03,SGBD94,SG95}. As observed in Fig.\ \ref{fig5},
the main influence of inelasticity is to decrease the value of the
critical shear rate.

\begin{figure}[tbp]
\includegraphics[width=.60 \columnwidth]{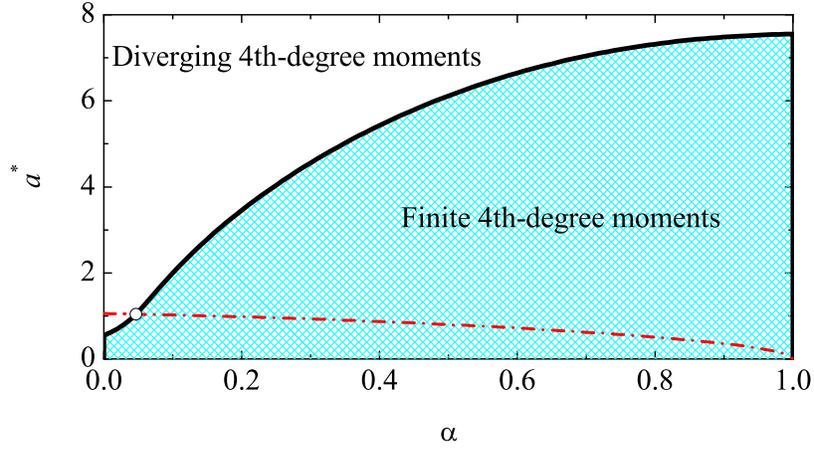}
\caption{(Color online) Phase diagram for the  asymptotic long time
behavior of the fourth-degree (symmetric) moments. The shaded region
below the curve $a_c^*(\alpha)$ (thick solid line) corresponds to
states with finite asymptotic values of the scaled fourth-degree
moments, while the region above the curve defines the states where
those moments diverge in time. The dash-dotted line represents the
steady-state points  $a_s^*(\alpha)$. It intercepts the critical
curve $a_c^*(\alpha)$ at the point $(\alpha,a^*)=(0.046,1.041)$.
\label{fig6}}
\end{figure}

Let us analyze  the phase diagram associated with the singular
behavior of the fourth-degree moments. This is shown in Fig.\
\ref{fig6}, where the curve $a_c^*(\alpha)$ splits the parameter
space into two regions: the region below the curve corresponds to
states $(\alpha,a^*)$ with finite asymptotic values of the scaled
fourth-degree moments (i.e., $\ellmin>0$), while the region above
the curve defines the states where those moments diverge in time
($\ellmin<0$). Figure \ref{fig6} also includes the locus of
steady-state points $(\alpha,a_s^*)$ [cf.\ Eq.\
\protect\eqref{W16}]. Below the latter curve the inelastic cooling
dominates over the viscous heating and so the temperature decreases
in time, while the opposite happens above it. It is apparent that
the curve $a_s^*(\alpha)$  lies inside the region
$a^*<a^*_c(\alpha)$, except for the small interval $\alpha\leq
0.046$ or, equivalently, $a_s^*(0.046)=1.041\leq a^*\leq
a_s^*(0)=1.054$. In conclusion, in order to find diverging moments,
one has to consider states with rather large values of the shear
rate at which the viscous heating  is much higher than the
collisional cooling.

\begin{figure}[tbp]
\includegraphics[width=.60 \columnwidth]{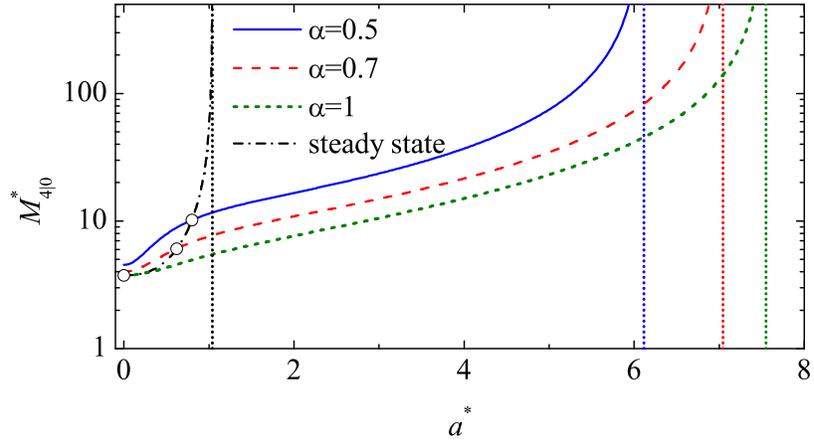}
\caption{(Color online) Plot of the asymptotic long time value of
the scaled moment $M_{4|0}^*$  as a function of $a^{\ast }$ for
$\alpha =0.5$ (solid line), $\alpha=0.7$ (dashed line), and $\alpha
=1$ (dotted line).
 The dash-dotted line represents the values of
$M_{4|0}^*$ at the steady states $a_s^*(\alpha)$ for
$0.046<\alpha\le 1$. It intercepts the curves representing
$M_{4|0}^*(a^*)$  at the points indicated by circles. The vertical
dotted lines are the asymptotes of the curves.\label{fig7}}
\end{figure}

For states with $a^*<a^*_c(\alpha)$ the scaled (symmetric)
fourth-degree moments reach well-defined finite values in the
asymptotic long time limit. {}From Eq.\ \eqref{3.4:d17} one has
\beq
\lim_{s\to\infty}\bm{\mathcal{M}}(s)=\bm{\mathcal{M}}_\infty,
\eeq
where $\bm{\mathcal{M}}_\infty$ is defined by Eq.\ \eqref{3.4:d18}.
As an illustration, Fig.\ \ref{fig7} shows the shear-rate dependence
of the asymptotic long time values of $M_{4|0}^*=(nv_0^4)^{-1}\int
\dd\mathbf{V} V^4 f(\mathbf{V})$ for $\alpha=0.5$, $0.7$, and $1$.
The values of $M_{4|0}^*$ at $a^*=0$ correspond to the homogeneous
cooling state \cite{GS07}. We observe that, given a value of
$\alpha$, the scaled moment $M_{4|0}^*$ rapidly increases with the
shear rate, having a vertical asymptote at $a^*=a_c^*(\al)$.
Moreover, for a given value of $a^*$, the value of the moment
increases with dissipation. Figure \ref{fig7} also includes the
curve representing the values of $M_{4|0}^*$ at the steady states
$a_s^*(\alpha)$. This curve has a vertical asymptote at $a^*=1.041$.

\section{Concluding remarks\label{sec5}}

The simple or uniform shear flow is perhaps the most widely studied
inhomogeneous state for elastic and inelastic gases. Despite its
apparent simplicity, this state has proven to be useful to shed
light on the nonlinear response of the system to the presence of
strong shearing. This response is conventionally measured by the
non-Newtonian rheological properties (namely, the nonlinear shear
viscosity $\eta$ and viscometric function $\Psi$), which are related
to the second-degree velocity moments (pressure tensor). On the
other hand, higher degree moments are also important to provide
indirect information on the features of the velocity distribution
function. For inelastic gases, there are two relevant control
parameters: the shear rate $a$ scaled with an effective collision
frequency $\nu_0$, i.e., the reduced shear rate $a^*=a/\nu_0$, and
the coefficient of normal restitution $\alpha$. When the velocity
moments are conveniently scaled with the thermal speed, they are
expected to become, after a kinetic transient regime, nonlinear
functions of both control parameters.

To address the above issues in the context of the Boltzmann equation
without having to resort to approximate methods or computer
simulations, one can consider simplified collision models. As in the
case of elastic collisions \cite{GS03}, the inelastic Maxwell model
(IMM) renders itself to an analytical treatment. Here, taking
advantage of a recent derivation by the authors of the collisional
moments for IMM through fourth degree \cite{GS07}, we have
determined the pressure tensor and the fourth-degree velocity
moments of the USF problem in an exact way. Two different classes of
IMM have been defined: Model A, where  $\nu_0$ is independent of
temperature, and Model B, where $\nu_0$ is an increasing function of
temperature. In the USF the temperature changes in time due to two
opposite effects: viscous heating and collisional cooling.
Therefore, a steady state is eventually achieved in Model B when
both effects cancel each other. However, in Model A a steady state
does not generally exist, except for a specific value
$a^*=a_s^*(\alpha)$, Eq.\ \eqref{W16}. It is important to note that
the results in the steady state are the same for both classes of
models. Since the reduced shear rate $a^*$  changes with time in
Model B, the goal of obtaining the velocity moments as functions of
$a^*$ and $\alpha$ requires the use of numerical tools in that case.
However, $a^*=\text{const}$ in Model A and so the independent
influence of $a^*$ and $\alpha$ can be studied analytically. Thus,
in this paper we have focused on Model A, except in what concerns
the steady state which, as said before, is common to Models A and B.

As mentioned above, the relevant transport properties in the USF
problem are the nonlinear shear viscosity $\eta^*(a^*)$ and the
viscometric function $\Psi(a^*)$. Their explicit forms are given by
Eqs.\ (\ref{W25}) and (\ref{W36}), respectively. These results
extend to inelastic collisions the expressions obtained long time
ago by Ikenberry and Truesdell for (elastic) Maxwell molecules
\cite{IT56}. With respect to the dependence of $\eta^*(a^*)$ and
$\Psi(a^*)$ on inelasticity, our results show that its influence  is
quite significant for small and moderate shear rates,  both
rheological properties being practically insensitive to dissipation
in the region of high shear rates. In the steady-state solution,
$\eta_s^*(a^*)$ and $\Psi_s(a^*)$ decrease when decreasing the
coefficient of restitution. Moreover, as expected, the (scaled)
third- and asymmetric fourth-degree moments vanish in the long time
limit. Consequently, beyond the rheological properties, the next
nontrivial moments are the symmetric fourth-degree moments. An
important result is that, for a given value of the coefficient of
restitution, these moments are divergent for shear rates larger than
a certain critical value $a_c^*(\alpha)$. This singular behavior is
also present in elastic systems \cite{GS03,SGBD94,SG95}, where it
has been shown that this divergence is consistent with an algebraic
high-velocity tail of the velocity distribution function. The main
effect of inelasticity is to decrease the value of $a_c^*(\alpha)$
as the gas becomes more inelastic. In addition, the phase diagram
associated with this singular behavior shows that the value of
$a_c^*(\alpha)$ is rather large in the whole domain $0<\alpha\leq
1$, so that in order to get diverging moments one has to consider
states at which the collisional cooling is strongly dominated by the
viscous heating effect. As a consequence,  nonlinear shearing
effects are still significant for $a^*<a_c^*$, as  illustrated in
Fig.\ \ref{fig7} for the scaled moment $M_{4|0}^*$.

The results derived in this paper can be useful for analyzing different situations. First, the knowledge of the
shear-rate dependence of the second- and fourth-degree moments of the USF allows one to determine the
generalized transport coefficients characterizing transport around the simple shear flow \cite{G07}. Another
possible direction of study is the extension of the present analysis for the rheological properties to
multicomponent systems. Previous works carried out for IMM \cite{G03,GA05} have shown the tractability of the
Maxwell kinetic theory for these complex systems and stimulate the performance of this study in the near future.

\acknowledgments

Partial support from the Ministerio de Educaci\'on y Ciencia (Spain)
through Grant No.\ FIS2007--60977 (partially financed by FEDER
funds) and from the Junta de Extremadura through Grant No.\ GRU07046
is acknowledged.

\end{document}